# Spin- and Voltage-Dependent Emission from Intra- and Intermolecular TADF OLEDs

Nikolai Bunzmann[1], Benjamin Krugmann[1], Sebastian Weissenseel[1], Liudmila Kudriashova[1], Khrystyna Ivaniuk[2], Pavlo Stakhira[2], Vladyslav Cherpak[2,3], Marian Chapran[4], Gintare Grybauskaite-Kaminskiene[5], Juozas Vidas Grazulevicius[5], Vladimir Dyakonov[1] and Andreas Sperlich[1*]

[1]Experimental Physics 6, Julius Maximilian University of Würzburg, 97074 Würzburg, Germany
[2]Department of Electronic Devices, Lviv Polytechnic National University, S. Bandera 12, 79013 Lviv, Ukraine
[3]Department of Physics, University of Colorado at Boulder, 390 UCB Boulder, CO 80309, USA
[4]Department of Molecular Physics, Lodz University of Technology, Zeromskiego 116, 90-924 Lodz, Poland
[5]Department of Polymer Chemistry and Technology, Kaunas University of Technology, Radvilenu pl. 19, LT-50254 Kaunas, Lithuania

**Abstract**

Organic light emitting diodes (OLEDs) based on thermally activated delayed fluorescence (TADF) utilize molecular systems with a small energy splitting between singlet and triplet states. This can either be realized in intramolecular charge transfer states of molecules with near-orthogonal donor and acceptor moieties or in intermolecular exciplex states formed between a suitable combination of individual donor and acceptor materials. Here, we investigate 4,4'-(9H,9'H-[3,3'-bicarbazole]-9,9'-diyl)bis(3-(trifluoromethyl) benzonitrile) (pCNBCzoCF$_3$), which shows intramolecular TADF but can also form exciplex states in combination with 4,4',4"-tris[phenyl(m-tolyl)amino]triphenylamine (m-MTDATA). Orange emitting exciplex-based OLEDs additionally generate a sky-blue emission from the intramolecular emitter with an intensity that can be voltage-controlled. We apply electroluminescence detected magnetic resonance (ELDMR) to study the thermally activated spin-dependent triplet to singlet up-conversion in operating devices. Thereby, we can investigate intermediate excited states involved in OLED operation and derive the corresponding activation energy for both, intra- and intermolecular based TADF. Furthermore, we give a lower estimate for the extent of the triplet wavefunction to be ≥ 1.2 nm. Photoluminescence detected magnetic resonance (PLDMR) reveals the population of molecular triplets in optically excited thin films. Overall, our findings allow us to draw a comprehensive picture of the spin-dependent emission from intra- and intermolecular TADF OLEDs.

*sperlich@physik.uni-wuerzburg.de

**I. Introduction**

Organic light emitting diodes (OLEDs) represent a promising alternative to conventional LEDs for display applications and room lighting. One of the major challenges in the development of OLED technologies has been the improvement of efficiencies since only 25% of injected charge carriers form emissive singlet states, while 75% form non-radiative triplet states (Brown1992, Rothberg1996, Bruetting2012). However, thermally activated delayed fluorescence (TADF) can be induced, if molecules exhibit a small energy splitting $\Delta E_{ST}$ between singlet and triplet states (Endo2009, Uoyama2012, Goushi2012). An enhanced reverse intersystem crossing (RISC) enables harvesting of triplets thus dramatically increases the efficiency of TADF based devices.

$\Delta E_{ST}$ is predominantly determined by the orbital overlap between the highest occupied molecular orbital (HOMO) and the lowest unoccupied molecular orbital (LUMO) of electrons and holes forming excitons (Endo2009). Therefore, the strategy to achieve small $\Delta E_{ST}$ is to design molecules which possess spatially separated HOMO and LUMO levels. This has either been achieved in molecules with twisted donor and acceptor moieties forming intramolecular charge transfer (CT) states (Dias2013, Zhang2014) or in suitable combinations of separate donor and acceptor molecules forming intermolecular exciplex states (Goushi2012, Hung2013, Li2014, Liu2015, Mamada2018). Both approaches have been implemented successfully and high external quantum efficiencies (EQE) in the range of 20% have been demonstrated (Zhang2014, Liu2015, Liu2016).

One of the remaining challenges is to tune the emission color of the used molecules for the desired application while maintaining high efficiency. An advantage of OLEDs for lighting applications is their surface-emission characteristic since conventional LEDs emit as potentially glaring point sources. Consequently, it is desirable to build efficient OLEDs which produce warm white light. The corresponding broad emission spectrum is often realized by blending the spectra of two single complementary emitters. One approach is the combination of a blue and an orange light source which has been realized by a blue fluorescent and an orange phosphorescent emitter (Schwartz2006, Qin2005). However, progress has also been made in producing white light based on TADF emitters (Zhao2015, Kaminskiene2018). One of these approaches is based on the molecule 4,4'-(9H,9'H-[3,3'-bicarbazole]-9,9'-diyl)bis(3-(trifluoromethyl) benzonitrile) (pCNBCzoCF$_3$) (Kaminskiene2018). It forms intramolecular CT states which exhibit sky-blue emission but additionally can form intermolecular exciplex states in combination with 4,4',4"-tris[phenyl(m-tolyl)amino]triphenylamine (m-MTDATA) which results in orange emission. A device combining both characteristics showed warm white emission color with high efficiency of 17.0% EQE (Kaminskiene2018).

Meanwhile, it is still an important point of discussion how the spin degree of freedom influences the first order forbidden RISC rate in operational TADF OLEDs. Since pCNBCzoCF$_3$ shows both intra- and intermolecular CT states, it is an interesting model system to employ inherently spin sensitive methods that just recently were applied for the first time to donor:acceptor based intermolecular TADF OLEDs (Vaeth2017, Bunzmann2020), but not yet to intra-molecular emitters or their combination. In this study, we investigate the TADF characteristics of the building blocks of warm white OLEDs based on the intramolecular CT emission from pristine pCNBCzoCF$_3$ and devices based on emission from exciplex states formed between pCNBCzoCF$_3$ and m-MTDATA. Moreover, we elucidate properties of the involved triplet states that are involved in the RISC mechanism for both device types by using electroluminescence detected magnetic resonance (ELDMR) for electrical generation and photoluminescence detected magnetic resonance (PLDMR) for optical excitation.

## II. Materials and Devices

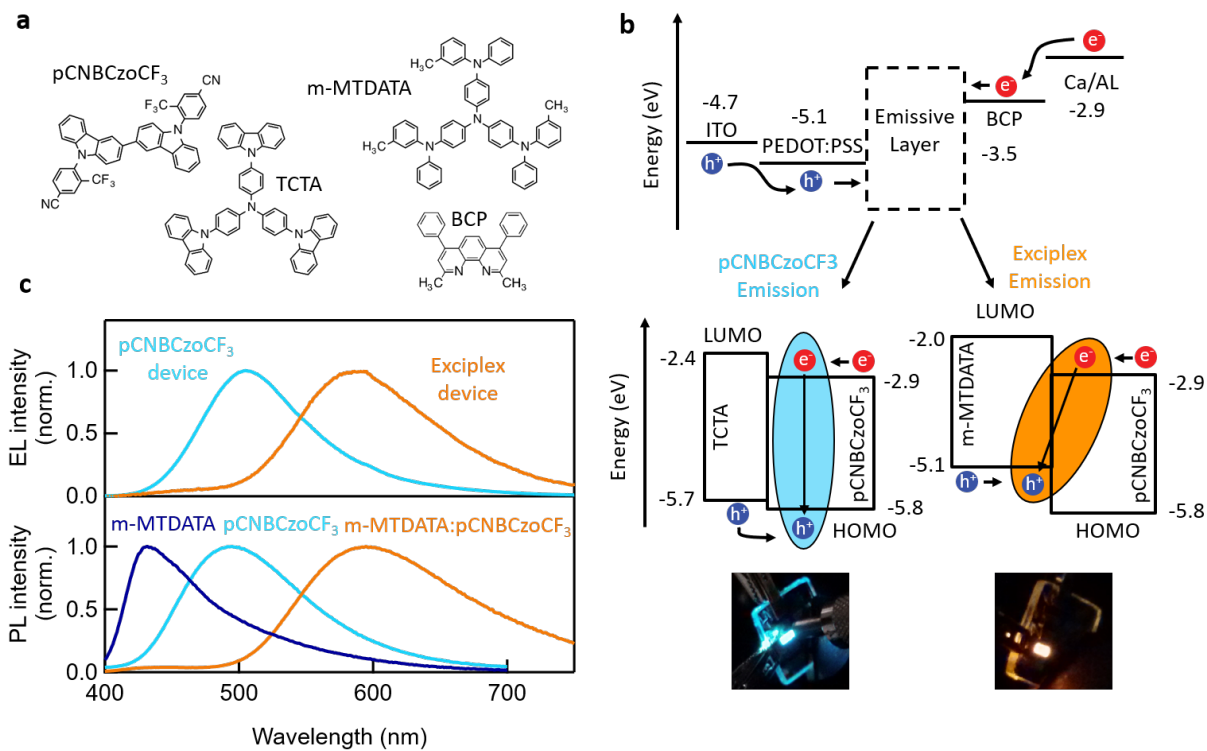

**Figure 1.** Materials and devices. **(a)** Chemical structures of molecules used in this work **(b)** Device structure and energy diagram of investigated OLEDs. Two different emissive layer combinations were used. For TCTA:pCNBCzoCF$_3$ there is no exciplex formation and emission originates from excitons on pCNBCzoCF$_3$. For m-MTDATA:pCNBCzoCF$_3$ emission originates from exciplex states at the interface. Photographs show the respective device under test. **(c)** EL spectra of both device types and corresponding PL spectra.

**Figure 1a** shows the molecules used in this work. pCNBCzoCF$_3$ and m-MTDATA are used for the emissive layers while TCTA (tris(4-carbazoyl-9-ylphenyl)amine) and BCP (2,9-dimethyl-4,7-diphenyl-1,10-phenanthroline) are used as transport layers. **Figure 1b** displays the corresponding device structure of OLEDs. Indium tin oxide (ITO) was used as an anode for all devices and Ca/Al as a cathode. Poly(3,4-ethylendioxythiophene):polystyrolsulfonate (PEDOT:PSS) and BCP were used as hole and electron transport layers respectively. Two types of devices were built with different choice of materials for the emissive layer. The first device type is based on TCTA and pCNBCzoCF$_3$ in the emissive layer (left part of **Figure 1b**). In this case, excitons form on pCNBCzoCF$_3$ giving rise to sky-blue emission as shown in the corresponding EL and PL spectra (**Figure 1c and S1**). Almost identical PL and EL spectra imply that there is no exciplex formation between TCTA and pCNBCzoCF$_3$. Instead, TCTA functions as an additional transport layer increasing the efficiency of the device. For the second type of device, the emissive layer consists of m-MTDATA and pCNBCzoCF$_3$ (right part of **Figure 1b**). Exemplary J-V-L curves for both device types are given in **Figures S2** and **S3**. Further device performance characteristics were reported previously by some of us and are discussed in Ref. (Kaminskiene2018).

For m-MTDATA:pCNBCzoCF$_3$ a clear redshift between the EL spectrum of the device and the PL spectra of pristine m-MTDATA or pCNBCzoCF$_3$ is recognizable (**Figure 1c**). The same shifted PL spectrum is observed for m-MTDATA:pCNBCzoCF$_3$ blend films which proves the formation of an exciplex state that gives rise to orange emission. There is a slight shoulder at 420-500 nm for both, PL and EL in **Figure 1c**, which can be assigned to additional emission from pCNBCzoCF$_3$ as shown in **Figure 2a**. To highlight the appearance of the shoulder, we subtracted the EL spectrum taken at 5 V from the EL spectrum at 9 V. The resulting spectrum (black) matches the PL spectrum of pristine pCNBCzoCF$_3$. Here, holes overcome the energetic barrier at the m-MTDATA:pCNBCzoCF$_3$ interface and form excitons with electrons in the pCNBCzoCF$_3$ layer, giving rise to the observed shoulder in the EL spectrum. This effect, however, does not need to be considered a deficiency since the use of an intramolecular TADF emitter as one of the constituents for an exciplex emitter can be beneficial for device efficiency (Liu2016). In such a case, triplets can undergo RISC on a TADF molecule, whereas they could be lost due to non-radiative decay on a conventional emitter molecule. As a result, a potential loss channel is averted. With a combination of the sky-blue emission of pristine pCNBCzoCF$_3$ and the orange emission of the exciplex, a warm white OLED can be realized (Kaminskiene2018, Sych2020).

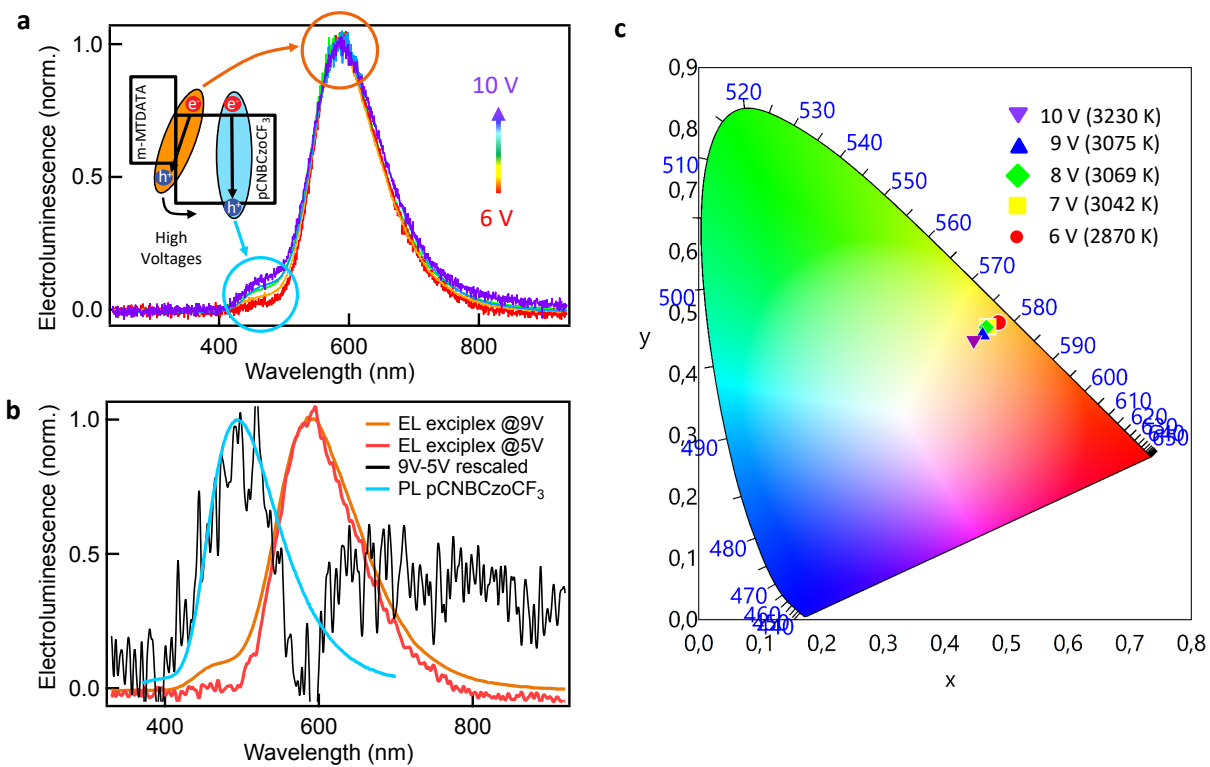

**Figure 2.** Color temperature tuning **(a)** Voltage dependent EL spectra of a device based on the exciplex between pCNBCzoCF$_3$ and m-MTDATA. There is a shoulder at 420-500 nm additionally to exciplex emission. The relative intensity of this shoulder is increasing with increasing voltage. **(b)** Comparison of a PL spectrum of pCNBCzoCF$_3$

and an EL spectrum of the exciplex-based device. The shoulder at 420-500 nm can be assigned to emission from pCNBCzoCF$_3$ by subtracting the normalized EL spectra at 9 V and 5 V (black). **(c)** CIE1976 chromaticity coordinates of the EL spectra presented in **(a)**. From 6 V to 10 V the color temperature shifts from 2870 K to 3230 K.

Remarkably, the relative intensity of the pCNBCzoCF$_3$ emission at the shoulder increases with increased driving voltage, as shown in **Figure 2a**. Between 6 V to 10 V the color temperature of the respective EL spectra shifts from 2870 K to 3230 K (**Figure 2c**). A device, that is optimized to exhibit emission from both pristine pCNBCzoCF$_3$ and the exciplex state, could be used to tune the emission in between warm and cold white by setting a different driving voltage. Meanwhile the brightness could be controlled independently by pulse width modulation. This concept is desirable as modern smartphone displays already employ a change in color temperature (so called "night shift") and this is also in demand for future room lighting applications. So far this can only be implemented by incorporating several OLEDs of different color and not within just one device.

If color purity is desired from the orange exciplex-based OLEDs, the corresponding devices can be realized with a 1:1 blended layer of m-MTDATA: pCNBCzoCF$_3$ in the emission layer. In this case the EL shoulder at 420-500 nm vanishes (**Figure S1**).

## III. Magnetic Resonance

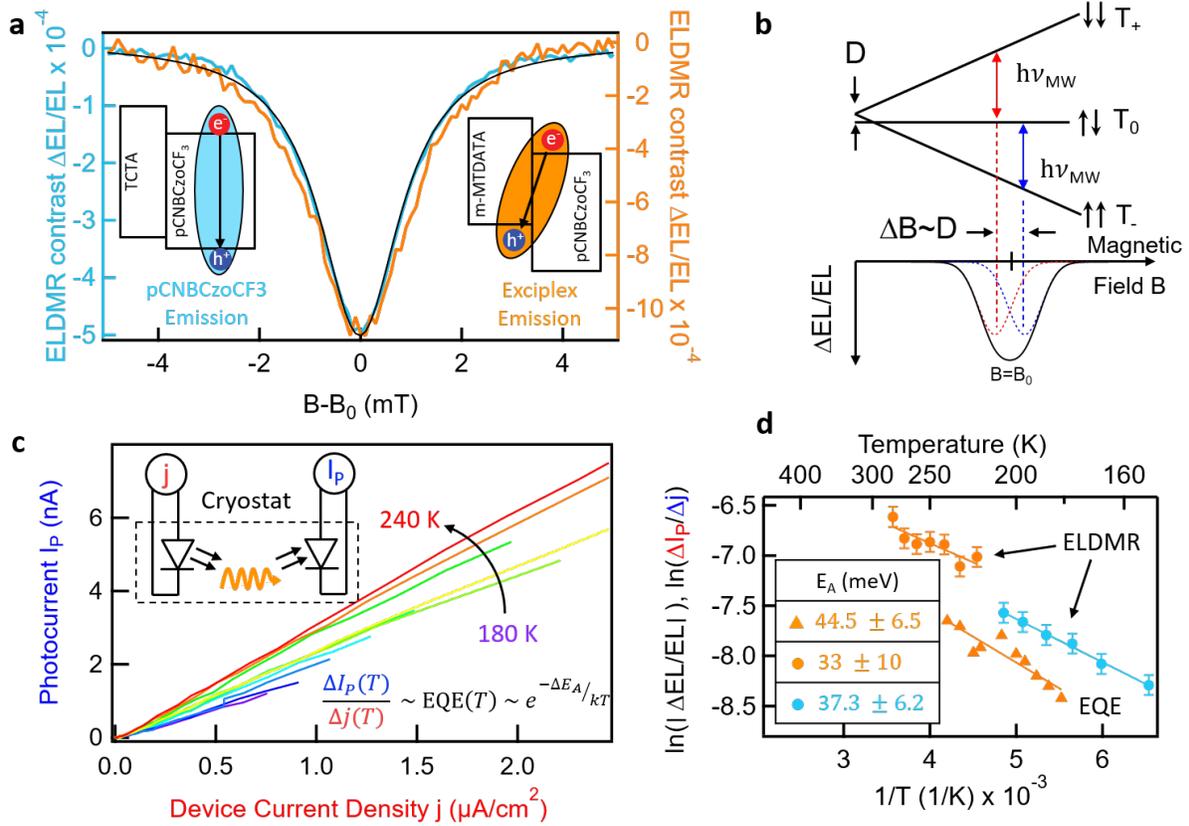

**Figure 3.** Magnetic resonance and activation energy. **(a)** Electroluminescence detected magnetic resonance (ELDMR) spectra of a pCNBCzoCF$_3$-based device and an exciplex-based device, recorded at 245 K. The magnetic field axis is shifted, such that resonance peaks at B=B$_0$ are centered around B-B$_0$. The scaling of left and right axis was chosen such that the shape of both spectra can be compared properly. A Lorentzian fit to the spectra is shown in black. **(b)** Illustration of the origin of the magnetic resonance spectra. The Zeeman energy levels of a CT triplet are split in an external magnetic field B and transitions that are induced by microwave photons ($\nu_{MW}$ = 8 GHz) can be detected in relative electroluminescence change ($\Delta$EL/EL) as one signal of two overlapping contributions centered at B=B$_0$ with its width determined by the dipolar interaction D. **(c)** OLED emission intensity (I$_P$) in dependence of device current density (j) for different temperatures. The inset shows a schematic illustration of the used cryostat setup. The OLED is driven by a current j while the emitted light induces a photocurrent I$_P$ in a photodetector placed in front of the OLED. The slope $\Delta$I$_P$/$\Delta$j is proportional to the EQE. For a TADF based device the EQE depends exponentially on the activation energy E$_A$ and on the inverse thermal energy 1/kT. **(d)** Arrhenius plot of the slopes of the I$_P$-j curves and the integrated ELDMR spectra. The activation energy E$_A$ is derived from the slope of linear fits for each data set.

Kaminskiene et al. showed the potential of pCNBCzoCF$_3$ for building warm white OLEDs (Kaminskiene2018). The focus of their work was to explore the spectral properties and performance characteristics of the corresponding devices. However, a detailed investigation of the TADF characteristics of pCNBCzoCF$_3$ as well as spin sensitive measurements, such as ELDMR, are still missing. These measurements elucidate properties of triplet states which are involved in the light generation mechanisms of the devices and the temperature dependence reveals the TADF activation energy (Vaeth2017, Bunzmann2020). The idea of ELDMR experiments is to probe EL, while applying a static magnetic field $B$ that induces a Zeeman splitting of triplet states. By applying resonant microwaves, the following resonance condition is fulfilled (Weil2007):

$$h\nu_{\text{MW}} = g\mu_B B \Delta m_s \pm D\left(\cos^2\theta - \frac{1}{3}\right) \tag{1}$$

Here $h$ is the Planck constant, $\nu_{\text{MW}}$ is the microwave frequency, $g = 2.002$ is the g-factor of the spin system, $\mu_B$ is the Bohr magneton, $\Delta m_s = 1$ is the allowed change of the magnetic quantum number, $D$ is the dipolar interaction and $\theta$ is the angle between the direction of the external magnetic field and the vector connecting the two spins of the triplet state. During an ELDMR measurement microwaves of fixed frequency $\nu_{\text{MW}} = 8$ GHz are applied while the magnetic field $B$ is swept. During this sweep the microwave-induced change of electroluminescence ΔEL is detected.

ELDMR spectra for OLEDs based on pCNBCzoCF$_3$ emission and on exciplex emission (1:1 mixed layer) are shown in **Figure 3a**. In both cases a change of EL arises in resonance. The intensity of the signals is different but the line shape is almost identical. The origin of these signals is illustrated in **Figure 3b**. Under resonant conditions rate equations that describe the interplay between triplet sublevels and the singlet state are altered, causing a change of the RISC rate and consequently a change in EL. The dipolar interaction $D$ lifts the degeneracy between triplet states with $m_s = 0$ and $m_s = \pm 1$ at $B=0$, which results in two resonant transitions at different magnetic fields. If $D$ is small, these two magnetic fields are very close. If the molecules are also randomly oriented the respective resonance curves overlap and appear as a single broadened peak. Accordingly, the corresponding linewidth is a measure for the strength of $D$. The exact value can, however, not solely be derived from the linewidth because other broadening mechanisms like unresolved hyperfine interactions with nearby nuclei are contributing as well. Instead, the full width at half maximum (FWHM) can be considered as an upper limit for $2D$. This estimation allows the calculation of a lower boundary for the extent of the triplet wave function, that is, the distance $r_{e-h}$ between electron and hole forming the triplet state. The following approximation can be made (Jeschke2002):

$$r_{e-h}[nm] = \sqrt[3]{\frac{2.785}{D[mT]}} nm \tag{2}$$

Here, one obtains $r_{e-h}$ in units of nm by using $D$ in units of mT. From the FWHM = 3 mT of the ELDMR spectra shown in **Figure 3a** one finds $2D \leq 3$ mT resulting in $r_{e-h} \geq 1.2$ nm. Such numbers fit to delocalized triplet states, where the electron-hole distance is large. In contrast, strongly localized molecular triplet excitons, that are being discussed to be involved in TADF emission (Dias2013, Santos2016), exhibit distinct broader spectra because of close electron-hole distance and thus strong dipolar interaction $D$ (Vaeth2016, Bunzmann2020). Such molecular triplets are, however, not observed in the studied material systems. Consequently, the narrow ELDMR linewidth is consistent with expectations for CT triplet states. For the device based on pure pCNBCzoCF$_3$ emission, this corresponds to the $^3$CT triplet, delocalized between the accepting and the donating moieties of pCNBCzoCF$_3$. For the exciplex-based device, it corresponds to the $^3$Exc triplet, formed between an electron located on pCNBCzoCF$_3$ and a hole located on m-MTDATA. The striking similarity of the two signals is evidence for almost identical wavefunction delocalization in both cases: a delocalized triplet is probed, that extents over different moieties of a single molecule or even two adjacent but separate molecules.

To investigate the TADF character of both device types, we measured temperature dependent ELDMR spectra (**Figure S5**). In both cases the signal shape stays identical, while the intensity decreases with decreasing temperature. This tendency is contrary to what is commonly observed in magnetic resonance experiments, where lower temperatures yield a higher signal intensity due to a higher spin polarization. Consequently, we assign the thermal response of the ELDMR spectra to the thermal activation of delayed fluorescence: $\Delta(EL/EL) \sim \exp(-E_A/k_BT)$. Due to this relation, temperature-dependent ELDMR can be used as a means to estimate the TADF activation energy independent of device characteristics (Vaeth2017, Bunzmann2020). A quantitative analysis of the signal intensities can be carried out via an Arrhenius plot, which allows us to derive an activation energy $E_A$ of the probed effect as shown in **Figure 3d**. For the device based on pure pCNBCzoCF$_3$ we find an activation energy of 37.3±6.2 meV and for the exciplex-based OLED 33±10 meV. Both device types yield $E_A$ in the range of thermal energy $k_BT$ which fits to TADF emitters. Hence, we consider these values to be an estimate for the singlet triplet splitting $\Delta E_{ST}$.

The determination of $\Delta E_{ST}$ for pCNBCzoCF$_3$ has shown considerable discrepancies in different previous reports. In (Kaminskiene2018) a $\Delta E_{ST}$ of 11 meV was reported, while (Cao2017) reported 190 meV. Both values were determined via the difference between emission peaks in PL and phosphorescence spectra of films at 77 K. This mismatch demonstrates that the determination of $\Delta E_{ST}$ via PL measurements is not unambiguous and has to be treated carefully. Our results support the value of (Kaminskiene2018), which is in the range of thermal energy at room temperature, in line with reasonable values for TADF emitters.

As a comparative method, we measured temperature-dependent EQE of the exciplex-based device (1:1 mixed layer) to corroborate our findings. The exact temperature-dependence of a TADF OLED's EQE is non-trivial (Tao2014, Gruene2020), however, in a first order approximation, one can assume: $EQE \sim \exp(-E_A/k_BT)$, with an activation energy $E_A$. A schematic illustration of the corresponding setup is shown in the inset of **Figure 3c**. An OLED is driven by a current j and the emitted light is collected with a photodiode placed in front of the OLED measuring the photocurrent $I_P$. Temperature dependent current density (j) and electroluminescence ($I_P$) to voltage (V) characteristics for an exciplex-based device are shown in **Figure S3**. In order to disentangle the effects of TADF and temperature dependent hopping transport in organic semiconductors, the photocurrent $I_P$ is plotted vs. the current density j in **Figure 3c**. The resulting curves yield a linear dependency. The slope $\Delta I_P/\Delta j$ of these lines is proportional to the EQE of the device. A quantitative analysis of the temperature dependence of these slopes is, therefore, equivalent to an analysis of the EQE itself. An Arrhenius plot allows the determination of the corresponding activation energy $E_A$, which is included in **Figure 3d**. Here an activation energy of 44.5±6.5 meV is determined for the exciplex based device, which is slightly higher than the value of 33±10 meV derived via ELDMR, but still consistent within the experimental errors. Overall, a value in the range of thermal energy is obtained, confirming the TADF character of the exciplex state between pCNBCzoCF$_3$ and m-MTDATA.

## IV. Photoluminescence Detected Magnetic Resonance

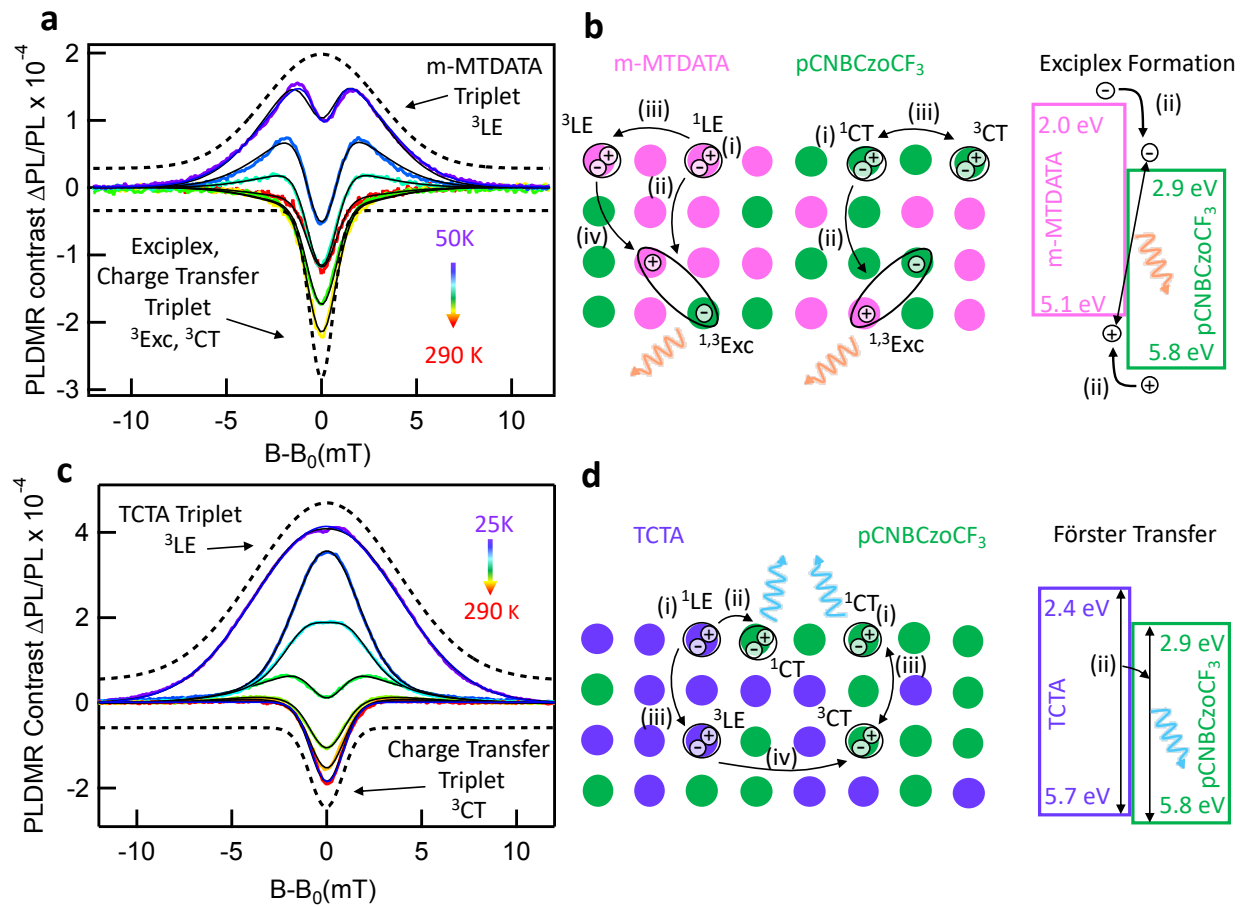

**Figure 4.** Magnetic resonance and exciton localization. Temperature dependent PLDMR spectra for blends of **(a)** m-MTDATA:pCNBCzoCF$_3$ and **(c)** TCTA:pCNBCzoCF$_3$ together with fits (black lines) of two superimposed Gaussian curves (black dashed lines). The narrow signal at higher temperatures is assigned to **(a)** the exciplex triplet or **(c)** the CT triplet on pCNBCzoCF$_3$. The broad signal at lower temperatures is assigned to **(a)** a local triplet on m-MTDATA or **(c)** TCTA. **(b,d)** Illustration of excitation pathways for singlet and triplet states in mixed blends. (i) Optical excitation of singlets on either m-MTDATA, TCTA or pCNBCzoCF$_3$. (ii) **(b)** Exciplex formation via CT from m-MTDATA to pCNBCzoCF$_3$. (ii) **(d)** Förster singlet transfer from TCTA to pCNBCzoCF$_3$. (iii) Formation of triplets on either m-MTDATA, TCTA or pCNBCzoCF$_3$ via ISC. (iv) **(b)** m-MTDATA triplets can undergo CT to exciplex triplets. (iv) **(d)** TCTA triplets can undergo Dexter transfer to pCNBCzoCF$_3$. Emissive states are $^1$CT on pCNBCzoCF$_3$ and singlet exciplexes. Temperature activated reverse ISC is possible for pCNBCzoCF$_3$ $^3$CT and triplet exciplex states enabling TADF. The right part of the figure displays the energies of HOMO and LUMO levels (Kaminskiene2018) with an additional illustration of interfacial exciplex formation (top) or Förster transfer (bottom).

We performed PLDMR measurements in order to explore the differences between the generation of excitons via optical excitation or electrical injection. In PLDMR, the external magnetic field is swept and the change of PL from an optically excited film is probed while resonant microwaves are applied.

First, we measured temperature-dependent PLDMR of a m-MTDATA:pCNBCzoCF$_3$ blend which is presented in **Figure 4a**. At low temperatures, there is a superposition of two signal components consisting of a broad positive signal and narrow negative signal. As temperature increases, the contribution of the broad signal vanishes and only the narrow one remains. The narrow component can be assigned to the CT triplet $^3$CT on pCNBCzoCF$_3$ and the exciplex triplet $^3$Exc, while the broad component can be assigned to the local triplet $^3$LE on m-MTDATA.

To explain the assignment of the observed signals, the consecutive steps leading to the population of different spin states are illustrated in **Figure 4b**. According to photoexcitation spectra of m-MTDATA and pCNBCzoCF$_3$ (**Figure S4a**) both, local singlets $^1$LE on m-MTDATA and CT singlets $^1$CT on pCNBCzoCF$_3$, are initially excited via the used 365 nm UV LED (i). The PL spectrum of a m-MTDATA:pCNBCzoCF$_3$ blend, however, shows a distinct redshift compared to the PL spectra of the pristine materials (**Figure S4a**). Consequently, a CT process leads to the formation of an exciplex state (ii). Here, either electron transfer from the LUMO of m-MTDATA to the LUMO of pCNBCzoCF$_3$ or hole transfer from the HOMO of pCNBCzoCF$_3$ to the HOMO of m-MTDATA takes place (see right side of **Figure 4b**). Exciplex singlets $^1$Exc decay radiatively or undergo ISC to form exciplex triplets $^3$Exc. Subsequent thermally activated RISC back to the singlet state induces delayed fluorescence. Combined prompt and delayed fluorescence give rise to the orange emission spectrum.

Alternatively to the CT process, ISC can facilitate the population of local triplets $^3$LE on m-MTDATA or CT triplets $^3$CT on pCNBCzoCF$_3$ (iii). Pristine m-MTDATA has a very low PL quantum efficiency (~7%) and can efficiently undergo ISC from optically excited $^1$LE to triplet excitons $^3$LE (Bunzmann2020). ISC on m-MTDATA and CT to pCNBCzoCF$_3$ are therefore competing processes. Electron and hole are delocalized over two molecules (exciplex) and for the pCNBCzoCF$_3$ excited states they are delocalized between donor:acceptor moieties of the same molecule. The distance between them is therefore relatively large, resulting in a small dipolar interaction $D$. As already discussed in **Figure 3b**, spin species with a small $D$ exhibit narrow signals in magnetic resonance experiments which is why the narrow component in PLDMR spectra of m-MTDATA:pCNBCzoCF$_3$ blends fits to the exciplex and the CT triplets. The broad signal can be assigned to the local triplet on m-MTDATA where electron and hole are localized on one molecule leading to a stronger dipolar interaction $D$ and consequently to a broader spectrum (Bunzmann2020).

The vanishing of the broad component at higher temperatures can be explained by the temperature dependencies of ISC on m-MTDATA, and the CT processes from m-MTDATA $^1$LE and $^3$LE to pCNBCzoCF$_3$. Charge transfer processes are mediated by molecular vibrations which are less pronounced at lower temperatures (Marcus1956). Consequently, the efficiency of the CT process decreases at low temperatures and the probability for ISC from m-MTDATA singlets to triplets increases which is why the broad signal is more pronounced at lower temperatures. Local m-MTDATA triplets can decay non-radiatively or, alternatively, still undergo CT to the exciplex triplet (iv) and thus contribute indirectly to delayed fluorescence from exciplexes. Consequently, spin manipulation of m-MTDATA triplets causes a change in exciplex emission intensity as observed by PLDMR, which enables indirect detection of non-emissive local triplet states.

In the next step, we measured PLDMR of a TCTA:pCNBCzoCF$_3$ blend which is presented in **Figure 4c**. Similarly, there is a superposition of a broad positive and a narrow signal where the broad component is predominantly pronounced at low temperatures and the narrow one at high temperatures. In this case the broad signal can be assigned to the local triplet $^3$LE on TCTA and the narrow one to the CT triplet $^3$CT on pCNBCzoCF$_3$.

To elaborate these assignments, the excitation pathways of spin states in an optically excited TCTA:pCNBCzoCF$_3$ blend are illustrated in **Figure 4d**. Via optical excitation either local singlets $^1$LE on TCTA or CT singlets $^1$CT on pCNBCzoCF$_3$ are generated (i) – as both materials can be excited at 365 nm according to photoexcitation spectra (see **Figure S4a** for pCNBCzoCF$_3$ and Ref. (Cao2018) for TCTA). The PL spectrum of a TCTA:pCNBCzoCF$_3$ blend exhibits only emission from pCNBCzoCF$_3$, while no emission from TCTA or red-shifted exciplex emission is observed (**Figure S4b**). Excitations on TCTA must therefore be depopulated efficiently. The absence of emissive exciplex states indicates that charge transfer is not expected. An alternative is singlet Förster transfer from TCTA to pCNBCzoCF$_3$ (ii).

Singlets on both materials can form local triplets $^3$LE on TCTA or CT triplets $^3$CT on pCNBCzoCF$_3$ via ISC (iii). The electron–hole separation of the local triplet on TCTA is small, whereas it is relatively large for the CT triplet on pCNBCzoCF$_3$. This means a strong dipolar interaction *D* for the local TCTA triplet and a relatively small *D* for the CT triplet. Therefore, the local TCTA triplet is observed as a broad signal and the CT triplet as a narrow signal.

As the TCTA triplet is detected by monitoring pCNBCzoCF$_3$ PL with PLDMR, subsequent Dexter triplet transfer between TCTA and pCNBCzoCF3 seems plausible (iv). Consequently, ISC on TCTA does not need to be considered as a loss mechanism, but just further delays emission.

## IV. Discussion

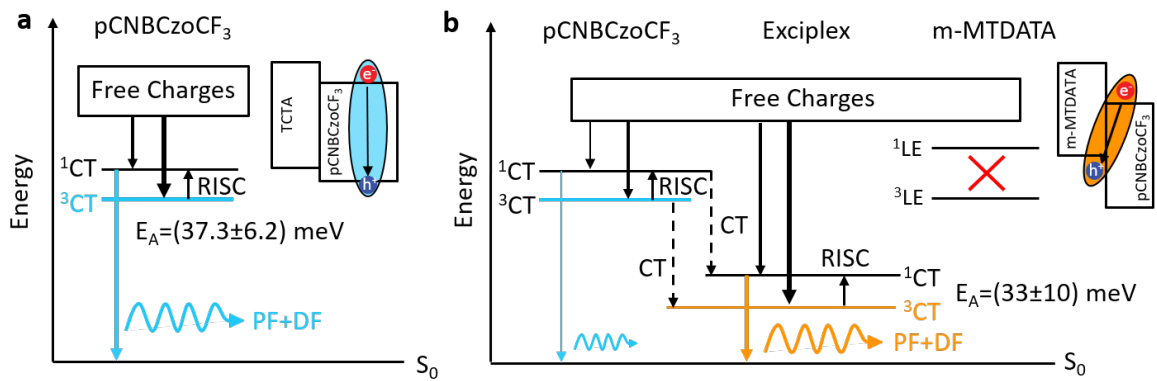

**Figure 5.** Energy diagrams of population pathways in electrically driven devices. **(a)** In the pCNBCzoCF$_3$-based device free charges populate CT singlets and triplets on pCNBCzoCF$_3$. Emission originates from the CT singlet via prompt fluorescence (PF) and via delayed fluorescence (DF) after RISC from the CT triplet to the CT singlet. **(b)** In the exciplex-based device, free charges populate singlets and triplets in the exciplex state but to some extent also on pCNBCzoCF$_3$. Emission originates from the exciplex singlet via prompt fluorescence and via delayed fluorescence after RISC from the exciplex triplet. Singlets on pCNBCzoCF$_3$ decay either radiatively or undergo CT to the exciplex singlet. Triplets on pCNBCzoCF$_3$ undergo RISC to form singlets or undergo CT to exciplex triplets. Activation energies of RISC processes are determined from temperature dependent ELDMR spectra. Neither is emission from m-MTDATA observed nor are signatures of local m-MTDATA triplets found in ELDMR spectra.

The relative positions of HOMO and LUMO levels in blends of pCNBCzoCF$_3$ with m-MTDATA or TCTA are similar. Yet, in one case an emissive exciplex state forms while in the other one only pCNBCzoCF$_3$ emits light. By performing both PLDMR and ELDMR, we were able to reveal the characteristic magnetic resonance signatures of triplet states in these systems and determined the respective activation energies for light emission. This information allows us to draw a comprehensive picture of the light generation mechanisms in pCNBCzoCF$_3$-based and in exciplex-based OLEDs.

The most important processes for the pCNBCzoCF$_3$-based device are summarized in **Figure 5a**. Injected charges populate CT singlet and CT triplet states with a ratio of 1:3. Singlets decay radiatively as prompt fluorescence while triplets undergo RISC giving rise to delayed fluorescence. Both processes yield the sky-blue emission spectrum of pCNBCzoCF$_3$. The triplet that is involved in RISC is found to be the CT triplet via ELDMR. Local triplets which are characterized by broad magnetic resonance spectra, as found in PLDMR, do not appear in ELDMR, and therefore do not play a role in the case of electrical injection. The activation energy of the RISC process is in the range of thermal energy proving that pCNBCzoCF$_3$ exhibits TADF.

All involved processes for the exciplex-based device are shown in **Figure 5b**. The majority of charges populates the singlet and triplet exciplex states. Singlets decay radiatively via prompt fluorescence while triplets undergo RISC giving rise to delayed fluorescence. These processes are responsible for the orange emission. The triplet which is involved in RISC was identified as the exciplex triplet via ELDMR. The local triplet of m-MTDATA was only found for optical excitation (PLMDR) but not in operational devices (ELDMR). The activation energy of the RISC process is close to thermal energy which shows that the exciplex exhibits TADF as well as pCNBCzoCF$_3$ itself. The EL spectrum of the exciplex-based device exhibits a shoulder which indicates additional population of CT singlets and triplets of pCNBCzoCF$_3$. Singlets can decay radiatively giving rise to the observed shoulder in the EL spectrum but can also undergo a CT to the exciplex singlet. Triplets on pCNBCzoCF$_3$ can undergo a CT to the exciplex triplet or undergo RISC to the CT singlet of pCNBCzoCF$_3$.

The direct comparison of excitation pathways for electrically driven OLEDs (**Figure 5**) and optically excited films (**Figure 4**) is intriguing. While the observed EL and PL emission spectra are independent of the type of excitation, different intermediate steps are observed by means of magnetic resonance. The reasons for this decisive difference are of course that only optical excitation directly excites singlet states on all involved materials, while the interfacial exciplex is not optically excitable. Injected charge carriers, however, recombine with a 75:25 triplet-to-singlet statistic primarily yielding interfacial exciplexes (m-MTDATA:pCNBCzoCF$_3$) or injected charges in pCNBCzoCF$_3$ recombine to form excitons. Consequently, neither m-MTDATA nor TCTA singlet $^1$LE and triplet $^3$LE excitons are involved for electrical excitation. Non-radiative decay of molecular excitons on m-MTDATA and TCTA as a loss mechanism is thus also a non-issue for electrical excitation in these or similar TADF systems. Consequently, this should be considered when using radiative quantum efficiency as a figure of merit to compare exciplex-based TADF systems.

**V. Conclusion**

The inherently spin-sensitive techniques of electroluminescence and photoluminescence detected magnetic resonance (ELDMR, PLDMR) were employed to study efficient OLEDs based on intra- and intermolecular TADF effects. From the results we can draw the following conclusions. The electrically generated intermediate triplet states that are responsible for light generation are broadly delocalized ($\geq 1.2$ nm) over either one molecule that features intramolecular TADF or over two adjacent molecules which exhibit exciplex type emission. Aside from that, no strongly localized triplet excitons were observed for electrically driven devices with ELDMR. However, upon optical excitation, such localized molecular triplet excitons on the donor materials are generated at low temperatures as evident by PLDMR. By performing temperature-dependent EQE and spin-resonance measurements on both types of devices, we derived singlet-triplet gaps $\Delta E_{ST}$ in the range of 33-45 meV which is in the order of thermal energy and hence in line with efficient TADF at ambient temperatures. Based on these studies, we propose an energy scheme of the recombination pathways for electrically driven OLEDs and optically excited films, highlighting that despite yielding the same emission spectrum, different intermediate steps are involved. Finally, we observed that for orange emitting exciplex-based devices a residual sky-blue emission from the used intra-molecular TADF emitter is generated, whose intensity can be voltage-controlled. This opens up the intriguing technical application of tuning both, the OLED emission color by voltage adjustments, while controlling its brightness by pulse width modulation.

**Experimental**

The materials m-MTDATA, TCTA and BCP were purchased from Sigma-Aldrich. pCNBCzoCF$_3$ was synthesized as described elsewhere (Kaminskiene2018). All materials were used as received.

All OLED devices were fabricated on indium tin oxide (ITO) covered glass substrates (1 cm$^2$). First, poly(3,4-ethylendioxythiophene):polystyrolsulfonate (PEDOT:PSS, 4083Ai) from Heraeus was spin coated with 3000 rpm for 1 minute, resulting in a 40 nm thick film. All further device fabrication steps were done inside a nitrogen glovebox to avoid degradation, starting with annealing of the 40 nm PEDOT:PSS layer for 10 minutes at 130°C. All further organic and metal contact layers were thermally evaporated in a vacuum chamber. For devices based on pristine pCNBCzoCF$_3$ the active layers were: TCTA (40 nm) / pCNBCzoCF$_3$ (40 nm). For exciplex bilayer devices: m-MTDATA (40 nm) / pCNBCzoCF$_3$ (40 nm). For exciplex devices with mixed emission layer for pure orange emission: m-MTDATA (30 nm) / 1:1 mixed emission layer (50 nm) / pCNBCzoCF$_3$ (30 nm). Finally, for all devices, BCP (10 nm) and the top electrode (5 nm Ca / 120 nm Al) were evaporated on top, completing 8 OLEDs (3 mm$^2$ each) per substrate.

For PL samples, the materials were evaporated onto glass substrates without the use of ITO, PEDOT:PSS, BCP or a metal cathode. PL and photoexcitation spectra were measured with a calibrated fluorescence spectrometer FLS980-s (Edinburgh Instruments) equipped with a continuous broad-spectrum xenon lamp Xe1.

EL spectra were recorded by biasing the OLED with an Agilent 4155C parameter analyzer in constant current mode and coupling the emitted light via light guides to an Acton Spectra SP-2356 spectrometer (Princeton Instruments) or a SPM002 spectrometer (Photon Control).

PLDMR samples were prepared from solutions of the materials in chlorobenzene. ~100 µl of the solutions were poured into EPR quartz tubes and the solvent was then evaporated by vacuum pumping. The sample tubes were subsequently sealed under inert helium atmosphere. Sample tubes were inserted into an EPR microwave cavity with optical access (Bruker ER4104OR) and an Oxford cryostat. Optical excitation was provided by a 365 nm UV LED behind a 400nm shortpass filter. The PL was detected by a silicon photodiode placed in front of the cavity behind a 409 nm longpass filter. Magnetic resonance measurements were done in two modified EPR spectrometers (Bruker E300, see **Figures S6, S7**) equipped with continuous flow helium cryostats (Oxford). OLED devices were placed in contact to a microwave transmission line and a silicon photodiode was mounted on top for detection of EL. Forward bias was provided by a source-measure unit (Keithley 237). EL, PL and bias currents were connected to current-voltage transimpedance amplifiers (Femto). The signal change upon resonant microwave irradiation was then detected via a Lock-In-Amplifier (SR7230) with the ≈kHz on-off modulated microwave as reference.

**Associated Content**
**Supporting Information**
J-V-L characteristics; PL, EL and ELDMR spectra; magnetic resonance setup schematics.
**Corresponding Author**
E-mail: sperlich@physik.uni-wuerzburg.de

**ORCID**

Benjamin Krugmann: 0000-0002-1879-7260

Sebastian Weissenseel: 0000-0001-9811-1005

Liudmila G. Kudriashova: 0000-0002-3793-5497

Khrystyna Ivaniuk: 0000-0003-1264-3532

Pavlo Stakhira: 0000-0001-5210-415X

Vladyslav Cherpak: 0000-0002-1879-4449

Marian Chapran: 0000-0002-2411-9642

Juozas Vidas Grazulevicius: 0000-0002-4408-9727

Vladimir Dyakonov: 0000-0001-8725-9573

Andreas Sperlich: 0000-0002-0850-6757

**Author Contributions**

The manuscript was written through contributions of all authors. All authors have given approval to the final version of the manuscript.

**Acknowledgements**

N.B. acknowledges support by the German Research Foundation, DFG, within the SPP 1601 (SP1563/1-1). S.W. acknowledges DFG FOR 1809 (DY18/12-2). M.C. acknowledges to the National Science Center, Poland (2018/31/N/ST4/03459). L.K, V.D. and A.S. acknowledge EU H2020-MSCA-ITN-2016 "SEPOMO" 722651. V.D. and A.S. acknowledge the German Research Foundation, DFG, within GRK 2112. J.V.G., K.I. and P.S. acknowledge EU H2020 research and innovation programme MSCA 823720. We thank Dr. G. Bagdziunas for advice in the synthesis.

**Notes**

The authors declare no competing financial interests

**Table of Contents Image**

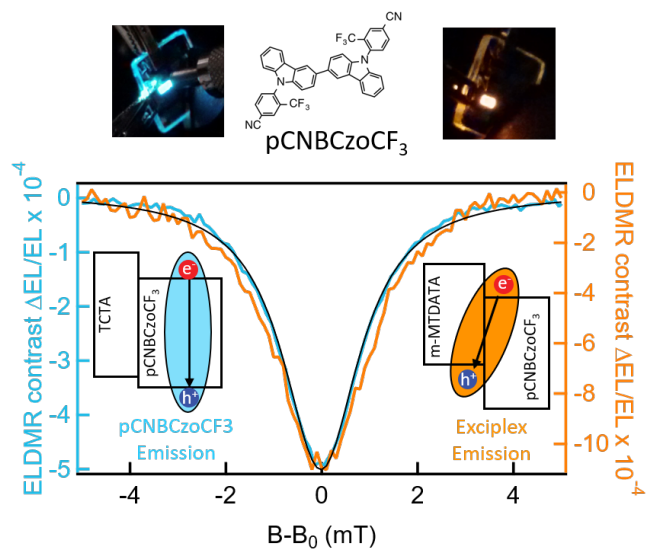

A molecular TADF emitter for sky blue OLEDs is studied. When combined with m-MTDATA, additional orange emission from interfacial exciplex states is observed which enables warm- to cold-white OLEDs with voltage-dependent color tuning. Excitation pathways for electrical generation and optical excitation are resolved using spin-sensitive magnetic resonance techniques.

Nikolai Bunzmann[1], Benjamin Krugmann[1], Sebastian Weissenseel[1], Liudmila Kudriashova[1], Khrystyna Ivaniuk[2], Pavlo Stakhira[2], Vladyslav Cherpak[2,3], Marian Chapran[4], Gintare Grybauskaite-Kaminskiene[5], Juozas Vidas Grazulevicius[5], Vladimir Dyakonov[1] and Andreas Sperlich[1*]

[1]Experimental Physics 6, Julius Maximilian University of Würzburg, 97074 Würzburg, Germany
[2]Department of Electronic Devices, Lviv Polytechnic National University, S. Bandera 12, 79013 Lviv, Ukraine
[3]Department of Physics, University of Colorado at Boulder, 390 UCB Boulder, CO 80309, USA
[4]Department of Molecular Physics, Lodz University of Technology, Zeromskiego 116, 90-924 Lodz, Poland
[5]Department of Polymer Chemistry and Technology, Kaunas University of Technology, Radvilenu pl. 19, LT-50254 Kaunas, Lithuania

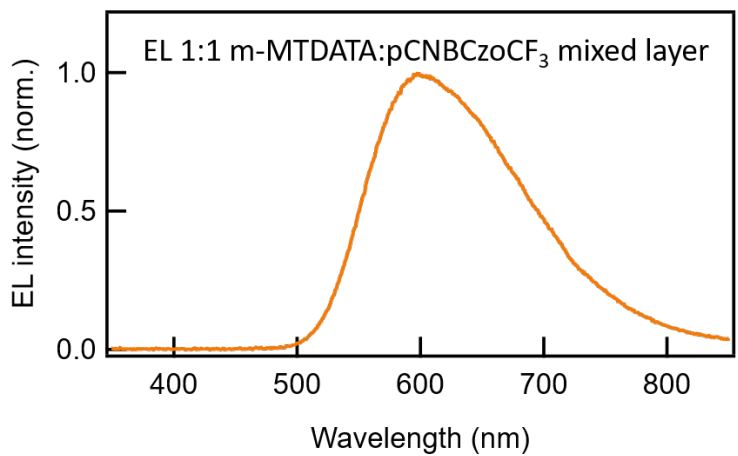

**Figure S1.** EL spectrum of an OLED with a 1:1 m-MTDATA:pCNBCzoCF$_3$ mixed layer in the emission layer showing pure exciplex emission.

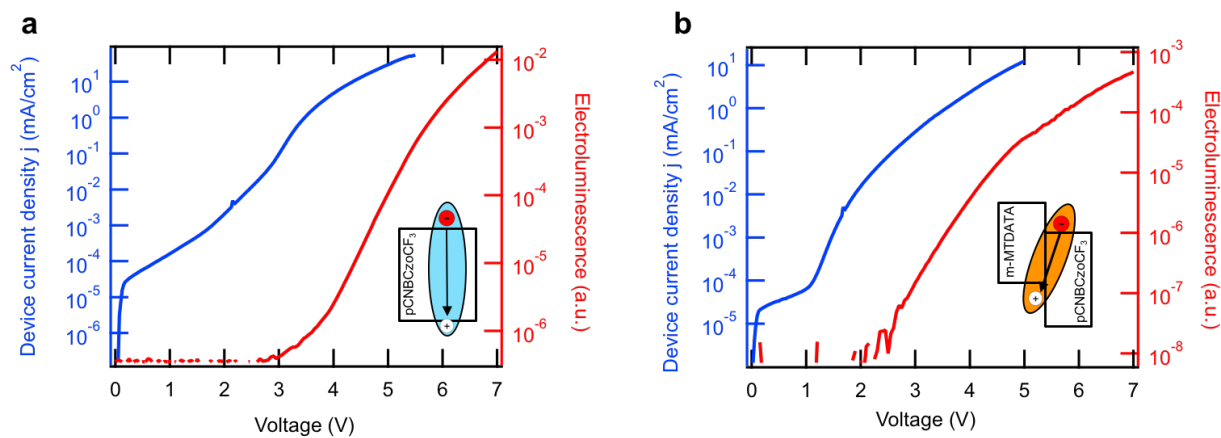

**Figure S2.** Current-density – voltage – electroluminescence curves for OLEDs with emission layers consisting of **(a)** TCTA:pCNBCzoCF$_3$ and **(b)** m-MTDATA:pCNBCzoCF$_3$.

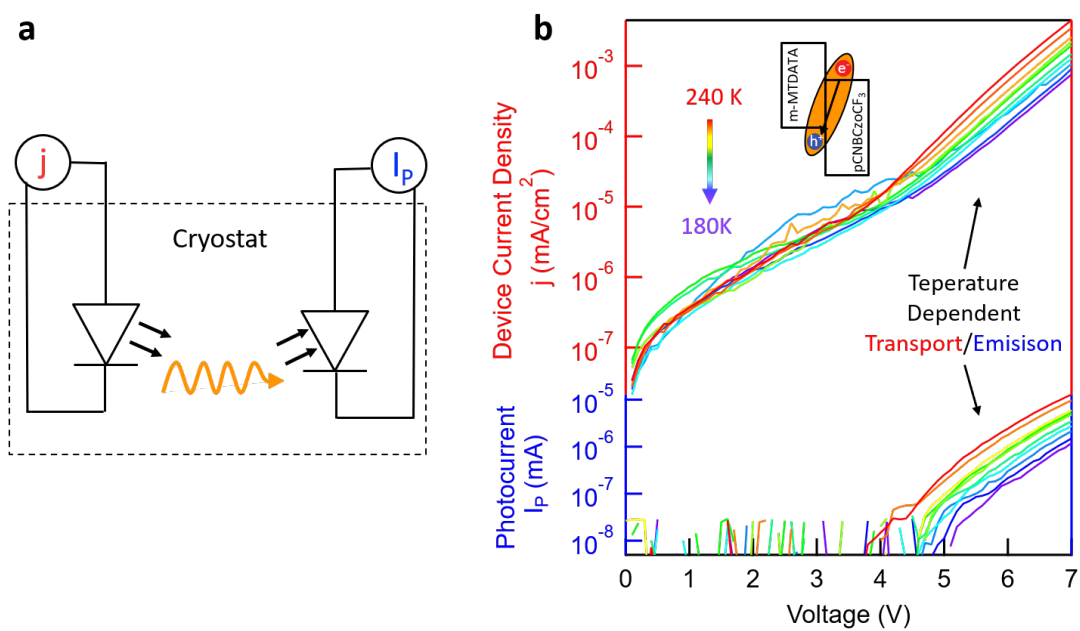

**Figure S3. (a)** Illustration of the setup used to measure temperature-dependent EQE. An OLED is driven by a current density j and the emitted light induces a photocurrent I$_P$ in a photodiode placed in front of the OLED. The setup is placed in a cryostat in order to measure at different temperatures. **(b)** Temperature-dependent current density and electroluminescence to voltage characteristics. Both the current density and the electroluminescence show a temperature activated regime.

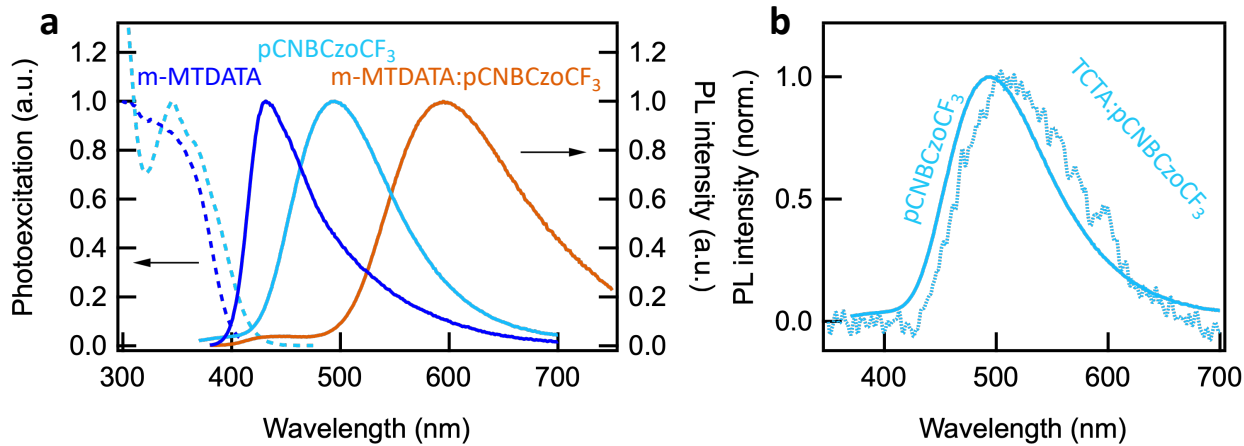

**Figure S4. (a)** Photoexcitation (PE) and PL spectra of m-MTDATA and pCNBCzoCF$_3$ and a PL spectrum of a m-MTDATA:pCNBCzoCF$_3$ blend. The PL spectrum of the blend shows that emission originates from the exciplex singlet. Excitation in PLDMR experiments is done by an UV-LED emitting at 365 nm. Therefore both, m-MTDATA and pCNBCzoCF$_3$, are excited according to the PE spectra. **(b)** PL spectra of pristine pCNBCzoCF$_3$ and blended TCTA:pCNBCzoCF$_3$. There is no exciplex formation between TCTA and pCNBCzoCF$_3$ as the PL of their blend is almost identical to pristine pCNBCzoCF$_3$.

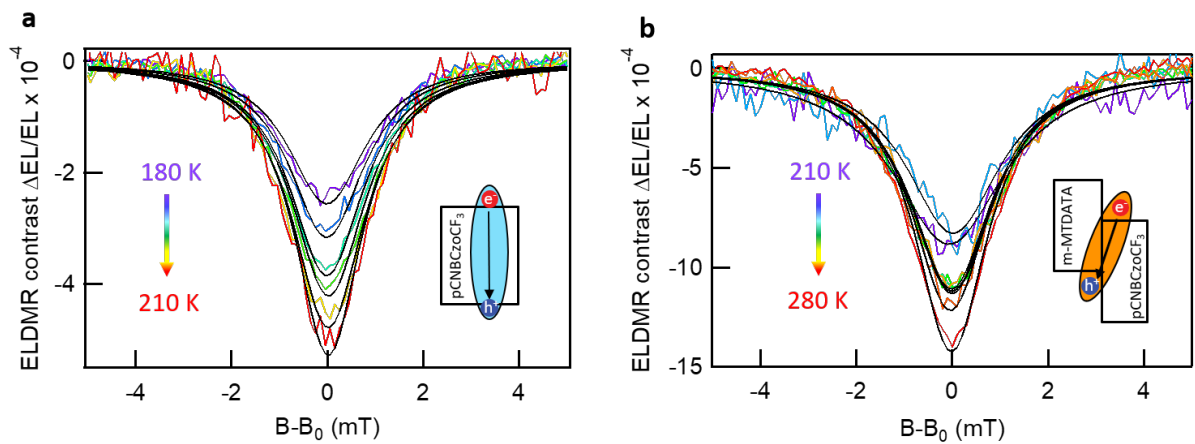

**Figure S5.** Temperature-dependent ELDMR spectra of **(a)** a pCNBCzoCF$_3$-based device and **(b)** an exciplex-based device together with Lorentzian fits (black lines). For both systems the signal decreases with decreasing temperature, which is in line with thermal activation of delayed fluorescence. Therefore, the temperature behavior shows that the observed effect is of TADF nature.

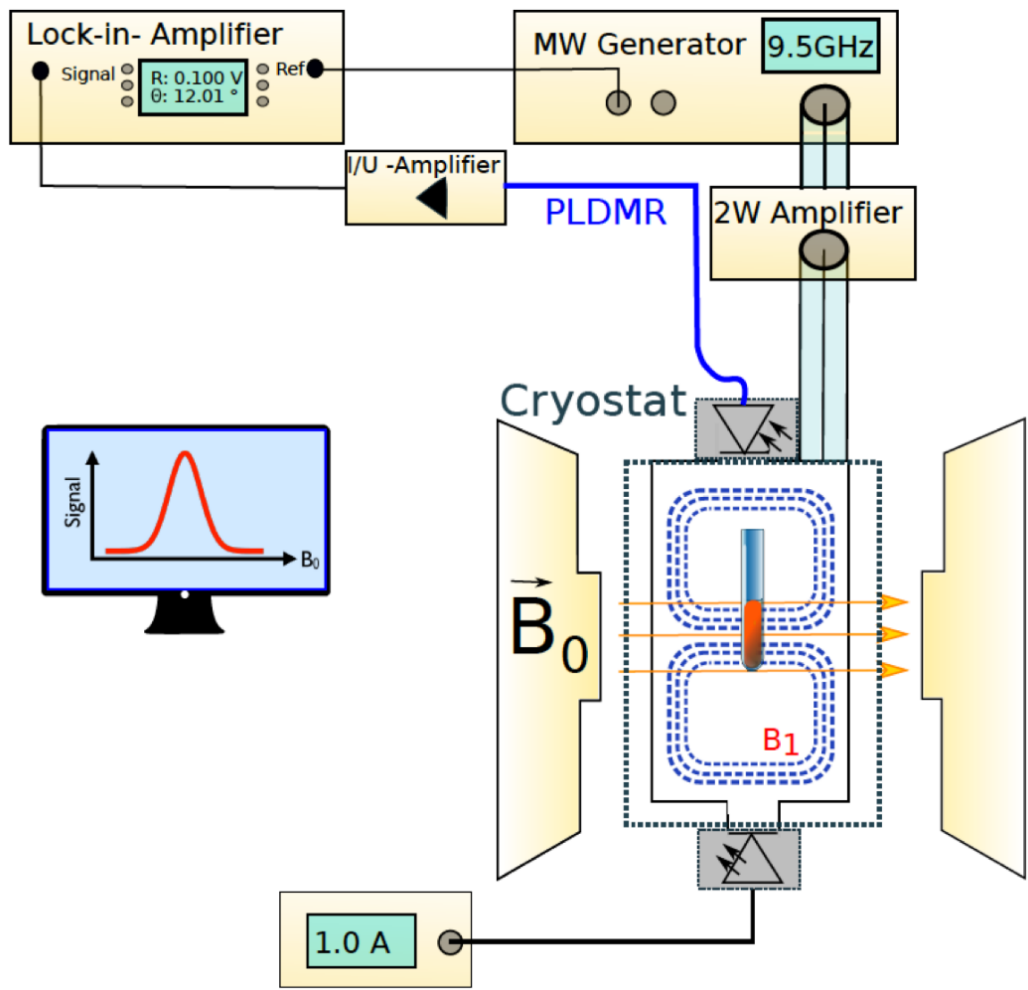

**Figure S6.** PLDMR setup.

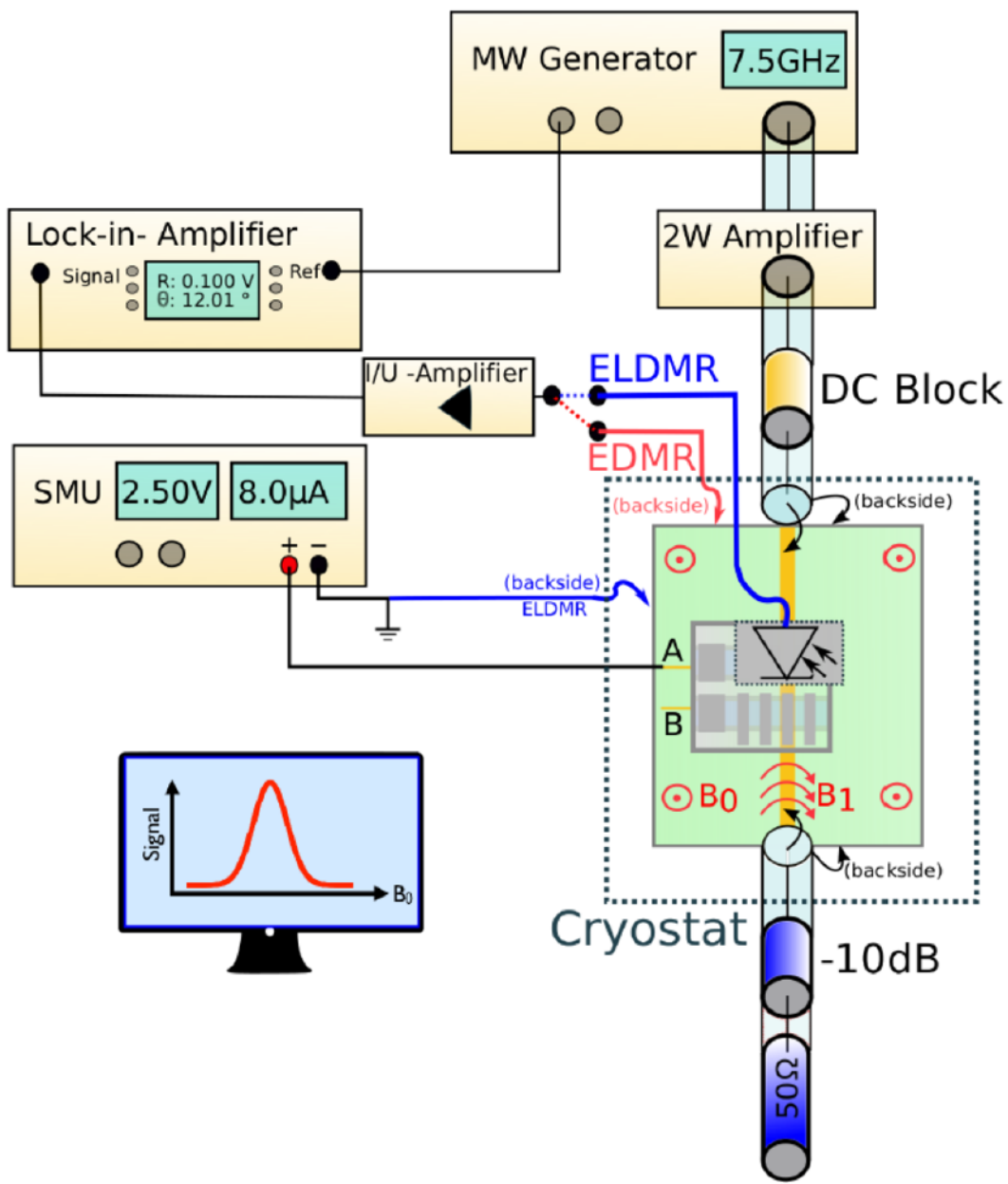

**Figure S7.** ELDMR setup.